\documentclass[review]{elsarticle}
\usepackage{lineno,hyperref,amsmath}

\modulolinenumbers[5]

\journal{Astroparticle Physics}

\begin{document} 

\begin{frontmatter}

\title{Chromatographic separation of radioactive noble gases from xenon}

\author[cwru,slac,kavli]{D.S.~Akerib} 

\author[icl]{H.M.~Ara\'{u}jo} 

\author[sdsmt]{X.~Bai} 

\author[icl]{A.J.~Bailey} 

\author[umd]{J.~Balajthy} 

\author[uedn]{P.~Beltrame} 

\author[yale]{E.P.~Bernard} 

\author[llnl]{A.~Bernstein} 

\author[cwru,slac,kavli]{T.P.~Biesiadzinski}

\author[yale]{E.M.~Boulton} 


\author[cwru,slac,kavli]{R.~Bramante}


\author[yale]{S.B.~Cahn} 

\author[ucsb]{M.C.~Carmona-Benitez} 

\author[brwn]{C.~Chan} 


\author[usd]{A.A.~Chiller} 

\author[usd]{C.~Chiller} 

\author[cwru]{T.~Coffey} 

\author[icl]{A.~Currie} 

\author[ucd]{J.E.~Cutter}  

\author[uedn]{T.J.R.~Davison} 


\author[lbnl]{A.~Dobi} 

\author[ucl]{J.E.Y.~Dobson} 

\author[uofr]{E.~Druszkiewicz} 

\author[yale]{B.N.~Edwards}

\author[lbnl]{C.H.~Faham} 

\author[brwn,lbnl]{S.~Fiorucci} 


\author[brwn]{R.J.~Gaitskell} 

\author[lbnl]{V.M.~Gehman} 

\author[ucl]{C.~Ghag} 

\author[cwru]{K.R.~Gibson} 

\author[lbnl]{M.G.D.~Gilchriese} 

\author[umd]{C.R.~Hall} 

\author[sdsmt,sdsta]{M.~Hanhardt} 

\author[ucsb]{S.J.~Haselschwardt}  

\author[ucb,yale]{S.A.~Hertel} 

\author[ucb]{D.P.~Hogan} 

\author[ucb,yale]{M.~Horn} 

\author[brwn]{D.Q.~Huang} 

\author[slac,kavli]{C.M.~Ignarra} 

\author[ucb]{M.~Ihm} 

\author[ucb]{R.G.~Jacobsen} 

\author[cwru,slac,kavli]{W.~Ji}

\author[ucb]{K.~Kamdin} 

\author[llnl]{K.~Kazkaz} 

\author[uofr]{D.~Khaitan} 

\author[umd]{R.~Knoche} 

\author[yale]{N.A.~Larsen} 

\author[cwru,slac,kavli]{C.~Lee\corref{cor1}} 

\author[ucd,llnl]{B.G.~Lenardo} 

\author[lbnl]{K.T.~Lesko} 

\author[lipc]{A.~Lindote} 

\author[lipc]{M.I.~Lopes} 


\author[ucd]{A.~Manalaysay} 

\author[tamu]{R.L.~Mannino} 

\author[uedn]{M.F.~Marzioni} 

\author[ucb,yale]{D.N.~McKinsey} 

\author[usd]{D.-M.~Mei} 

\author[ucd]{J.~Mock} 

\author[uofr]{M.~Moongweluwan} 

\author[ucd]{J.A.~Morad} 

\author[uedn]{A.St.J.~Murphy} 

\author[ucsb]{C.~Nehrkorn} 

\author[ucsb]{H.N.~Nelson} 

\author[lipc]{F.~Neves} 

\author[ucb,lbnl,yale]{K.~O'Sullivan} 

\author[ucb]{K.C.~Oliver-Mallory} 


\author[wisc,slac,kavli]{K.J.~Palladino} 



\author[yale]{E.K.~Pease} 

\author[cwru]{K.~Pech} 

\author[cwru]{P.~Phelps} 

\author[ucl]{L.~Reichhart} 

\author[brwn]{C.~Rhyne} 

\author[ucl]{S.~Shaw} 

\author[cwru,slac,kavli]{T.A.~Shutt} 

\author[lipc]{C.~Silva} 

\author[lipc]{V.N.~Solovov} 

\author[lbnl]{P.~Sorensen} 

\author[ucd]{S.~Stephenson}  

\author[icl]{T.J.~Sumner} 

\author[ualbany]{M.~Szydagis} 

\author[sdsta]{D.J.~Taylor} 

\author[brwn]{W.~Taylor} 

\author[yale]{B.P.~Tennyson} 

\author[tamu]{P.A.~Terman} 

\author[sdsmt]{D.R.~Tiedt}

\author[cwru,slac,kavli]{W.H.~To} 

\author[ucd]{M.~Tripathi} 

\author[yale]{L.~Tvrznikova} 

\author[ucd]{S.~Uvarov} 

\author[brwn]{J.R.~Verbus} 


\author[tamu]{R.C.~Webb} 

\author[tamu]{J.T.~White}

\author[cwru,slac,kavli]{T.J.~Whitis}

\author[ucsb]{M.S.~Witherell} 

\author[uofr]{F.L.H.~Wolfs} 


\author[icl]{K.~Yazdani} 

\author[ualbany]{S.K.~Young} 

\author[usd]{C.~Zhang} 

\cortext[cor1]{Corresponding Author: cxl370@case.edu; Present address: Center for Underground Physics, Institute for Basic science (IBS), Daejeon 305-811, Republic of Korea}

\address[cwru]{Case Western Reserve University, Dept. of Physics, 10900 Euclid Ave, Cleveland OH 44106, USA}
\address[slac]{SLAC National Accelerator Laboratory, 2575 Sand Hill Road, Menlo Park, CA 94025}
\address[kavli]{Kavli Institute for Particle Astrophysics and Cosmology, Stanford University, 452 Lomita Mall, Stanford, CA 94309, USA}
\address[icl]{Imperial College London, High Energy Physics, Blackett Laboratory, London SW7 2BZ, UK}
\address[sdsmt]{South Dakota School of Mines and Technology, 501 East St Joseph St., Rapid City SD 57701, USA}
\address[umd]{University of Maryland, Dept. of Physics, College Park MD 20742, USA}
\address[uedn]{SUPA, School of Physics and Astronomy, University of Edinburgh, Edinburgh EH9 3FD, United Kingdom}
\address[yale]{Yale University, Dept. of Physics, 217 Prospect St., New Haven CT 06511, USA}
\address[llnl]{Lawrence Livermore National Laboratory, 7000 East Ave., Livermore CA 94550, USA}
\address[ucsb]{University of California Santa Barbara, Dept. of Physics, Santa Barbara, CA, USA}
\address[brwn]{Brown University, Dept. of Physics, 182 Hope St., Providence RI 02912, USA}
\address[usd]{University of South Dakota, Dept. of Physics, 414E Clark St., Vermillion SD 57069, USA}
\address[ucd]{University of California Davis, Dept. of Physics, One Shields Ave., Davis CA 95616, USA}
\address[lbnl]{Lawrence Berkeley National Laboratory, 1 Cyclotron Rd., Berkeley CA 94720, USA}
\address[ucl]{Department of Physics and Astronomy, University College London, Gower Street, London WC1E 6BT, United Kingdom}
\address[uofr]{University of Rochester, Dept. of Physics and Astronomy, Rochester NY 14627, USA}
\address[sdsta]{South Dakota Science and Technology Authority, Sanford Underground Research Facility, Lead, SD 57754, USA}
\address[ucb]{University of California Berkeley, Department of Physics, Berkeley CA 94720, USA}
\address[lipc]{LIP-Coimbra, Department of Physics, University of Coimbra, Rua Larga, 3004-516 Coimbra, Portugal}
\address[tamu]{Texas A \& M University, Dept. of Physics, College Station TX 77843, USA}
\address[ualbany]{University at Albany, State University of New York, Department of Physics, 1400 Washington Ave., Albany, NY 12222, USA}
\address[wisc]{University of Wisconsin-Madison, Department of Physics, 1150 University Ave., Madison, WI 53706, USA}

\begin{abstract}
The Large Underground Xenon (LUX) experiment operates at the Sanford Underground Research Facility to detect nuclear recoils from the hypothetical Weakly Interacting Massive Particles (WIMPs) on a liquid xenon target. Liquid xenon typically contains trace amounts of the noble radioactive isotopes $^{85}$Kr and $^{39}$Ar that are not removed by the {\em in situ} gas purification system. The decays of these isotopes at concentrations typical of research-grade xenon would be a dominant background for a WIMP search experiment. To remove these impurities from the liquid xenon, a chromatographic separation system based on adsorption on activated charcoal was built. 400\,kg of xenon was processed, reducing the average concentration of krypton from 130\,ppb to 3.5\,ppt as measured by a cold-trap assisted mass spectroscopy system. A 50 kg batch spiked to 0.001 g/g of krypton was processed twice and reduced to an upper limit of 0.2 ppt.
\end{abstract}


\end{frontmatter}


\section{Introduction}

Liquid xenon is an excellent target for the direct detection of WIMP dark matter~\cite{Chepel:2013fy}, particularly when instrumented in a time projection chamber (TPC) as in the LUX detector~\cite{Akerib:2014uk}. Xenon's high proton number allows a very short penetration depth of external gamma and beta radiation~\cite{NISTSTAR,NISTXCOM}, and the event-by-event position measurement of the TPC allows these backgrounds to be highly suppressed in the inner volume of the detector. Because it has no long-lived radioactive isotopes~\cite{Firestone:471274}, xenon is intrinsically  quiet. The average single-scatter rate in the energy window of 0.9--5.3~keV$_\text{ee}$\footnote{The energy window calibrated with electronic recoils.} inside the 118~kg fiducial mass in LUX is measured to be below $10^{-3}$ events per kg/day/keV (differential rate unit, DRU$_{\text{ee}}$)~\cite{LUXBG2014}. This rate is dominated by the gamma rays from radioactive impurities in the 122 Hamamatsu R8778 photomultiplier tubes (PMTs)~\cite{Akerib:2013wu}.

Xenon, being distilled from the atmosphere, contains noble radioactive impurities such as $^{85}$Kr and $^{39}$Ar with half-lives of 10.756~yrs and 269~yrs, respectively.~\cite{Kr85Basics, WARPArgon}. Their characteristics are summarized in Table~\ref{tb:krAndArinXe}. 

$^{85}$Kr is generated by anthropogenic fission, and released into the atmosphere primarily during nuclear fuel reprocessing~\cite{Ahlswede:2013kv}. It contributes about 1~Bq/m$^3$ of the radioactivity from atmosphere~\cite{Cauwels:2001ky}, from which one can deduce that about 10~parts-per-trillion (ppt, $10^{-12}$) (g/g)\footnote{Concentrations are quoted by the ratio of masses unless quoted otherwise. Parts per million~($10^{-6}$), billion~($10^{-9}$), trillion~($10^{-12}$), and quadrillion~($10^{-15}$) are abbreviated as ppm, ppb, ppt, and ppq.} of atmospheric Kr is $^{85}$Kr. A measurement based on low-level counting reported 4--22.5~ppt~\cite{Du:dx}. Research-grade xenon contains about $10^{-7}$ natural krypton by mass. One analysis of boil-off gas of a distillation tower revealed that $6\pm2$~ppt of the krypton impurity is $^{85}$Kr in their sample~\cite{Abe:2013by}. At these concentrations, the decay of $^{85}$Kr yields a rate of about 6~DRU$_\text{ee}$, which overwhelms the potential dark matter signal. To be comparable to the background rate due to the PMTs, the xenon in LUX must contain less than 20~ppt krypton. The goal for LUX to reduce the $^{85}$Kr concentration to below 4~ppt was met using the method described in this paper. 

$^{39}$Ar is mainly generated by the $^{40}\text{Ar}+\text{n}\rightarrow\,^{39}\text{Ar}+2\text{n}$ process in the atmosphere due to cosmic rays, and constitutes about 1~ppq of argon in the atmosphere~\cite{WARPArgon}. The isotope emits a beta particle with an endpoint of 0.565~MeV. The LUX research-grade xenon originally contained about 1~ppb of argon. However, a portion of xenon was retrieved from other experiments, and contained an unknown amount of argon. For its background rate to be comparable to that from the photomultiplier tubes in the fiducial volume, the argon concentration must be reduced below ppb. 

\begin{table}
\caption{Characteristics of Radioactive $^{85}$Kr and $^{39}$Ar in Xenon.}
\begin{center}
\begin{tabular}{c | c | c}
 & $^{85}$Kr~\cite{Kr85Basics} & $^{39}$Ar~\cite{WARPArgon} \\
\hline
decay mode & $\beta^{-}$ & $\beta^{-}$ \\
half-life (years) & 10.756 & 269 \\
Q-value (keV) & 687 & 565 \\
abundance of the radioactive isotope (g/g) & $6\pm2$~ppt & 1~ppq \\
elemental concentration in original xenon (g/g) & $0.13$~ppm & unknown \\
elemental concentration goal (g/g) & $<4$~ppt & $<1$~ppb \\
\end{tabular}\label{tb:krAndArinXe}
\end{center}
\end{table}

Cryogenic distillation has been used to separate these light radioactive noble impurities from xenon. The XMASS detector is a single-phase dark matter detector containing 800~kg of liquid xenon. The XMASS collaboration developed a cryogenic distillation column to reduce the krypton level in xenon. Their distillation column lowered the krypton concentration by a factor of $1,000$ to 1.9~ppt of krypton~\cite{XMASSKr}. A similar system was adopted by the XENON100 collaboration~\cite{Aprile:2012bj}, and dropped their krypton concentration below 1~ppt~\cite{Lindemann:2014ea}.

\section{Adsorption-Based Chromatography} 
Adsorption-based chromatography is widely used for gas separation in industrial and chemical applications. A common industrial application is the separation of nitrogen from air, known as ``pressure swing adsorption." Among the scientific applications, the Borexino~\cite{BorexinoRd} and NEMO-2~\cite{Nachab:2006wt} collaborations developed a charcoal adsorption system to remove atmospheric radon from underground laboratory air, while a similar system removed krypton from xenon for the XENON-10 experiment~\cite{ABolozKr}. The latter system processed 20~kg of xenon in 2~months, with the final krypton concentration below 3~ppt. In this section, we present a mathematical treatment of adsorption-based gas chromatography central to our application, following the approach presented in \cite{BorexinoRd}. More general reviews are available in the literature~\cite{Engewald:2014jr}.

Adsorption refers to adhesion of atoms or molecules on a surface. The typical binding energy for adsorption is smaller than that of covalent bonds, and the process is reversible:
\begin{equation}\label{eq:adsorption}
	R + X \rightleftharpoons RX.
\end{equation}
Here, $X$ is the molecule to be adsorbed, or adsorbate, on the sites provided by $R$, the adsorbent. 

The equilibrium between the free and the adsorbed states of Eq.~\ref{eq:adsorption} depends principally on a few parameters:  the adsorbate, the adsorbent, the ambient temperature, and the concentration of adsorbate. A simple parametrization for mono-layer adsorption was proposed by Langmuir~\cite{Langmuir:1918fw}. The fraction $\theta$ of the sites occupied is described as a function of the partial pressure $P$ of the adsorbent:
\begin{equation}\label{eq:langmuir}
	\theta = \frac{KP}{1+KP}.
\end{equation}
$K$ is the equilibrium constant, the ratio of adsorption and desorption rates in Eq.~\ref{eq:adsorption}. It has a dimension of inverse pressure because the rate of adsorption is proportional to the partial pressure of $X$. If $P$ is large, $\theta$ converges to 1, indicating that all sites are occupied.

At low $P$ ($KP\ll 1$), $\theta$ depends linearly on $P$:
\begin{equation}\label{eq:henry}
	S = S_0 \theta = S_0 K P = b P.
\end{equation}
Here, $S$ is the molecular density of the adsorbate on the adsorbed state per unit mass of adsorbent, and $S_0$ is the maximum value of $S$. Therefore, $S$ is proportional to the partial pressure of the adsorbate in the free state, and the new constant $b$ is referred to as the Henry constant.

The partial pressure $P$ is proportional to the number density $y=\frac{N}{V}$ of the adsorbate in the free state by the ideal gas law, which allows us to write:
\begin{equation} 
S =  b k_BT y = ky	\label{eq:adsorptionConstant}.
\end{equation}  
The new constant, $k$, is called the adsorption constant, and is a ratio of the molecular densities of the adsorbate on the adsorbent ($N_{ads}/M$) to that of the gas phase ($N_g/V$): 
\begin{equation}
k = \frac{N_{ads}/M}{N_g/V}.
\end{equation} 
It quantifies the affinity of the adsorbate to the adsorbent. Differences in the adsorption constant (or the Henry constant) between adsorbates leads to different duration spent by the adsorbates in the desorbed phase. For example, the adsorption constants of xenon and krypton, calculated from \cite{Munakata99} at 300~K, are 1.15 and 0.064~l/g, respectively, different by a factor of 19.

\begin{figure}
\centering
\includegraphics[width=0.7\textwidth]{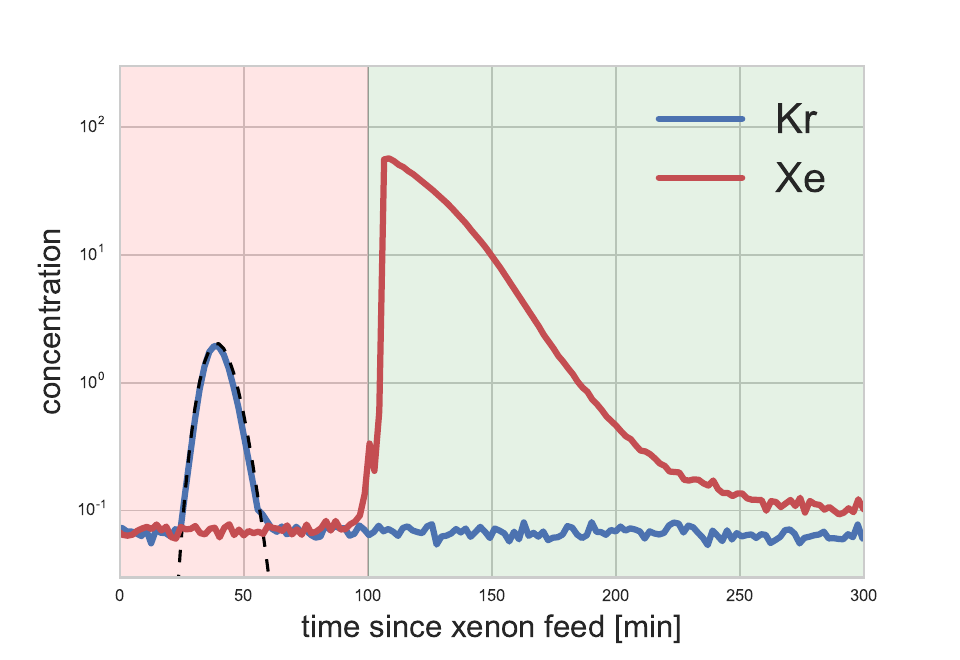}
\caption[Separation example]{Gas mixture separated by a chromatographic column. The $y$-axis represents the concentration in arbitrary units as measured with an residual gas analyzer~(RGA). The xenon is mixed with $10^{-2}$ mole fraction krypton to enhance the krypton concentration above the RGA baseline. The $x$-axis is  minutes since the beginning of the xenon feed, which lasts for the first 15 minutes. During the first 100~minutes, the helium-krypton mixture exiting the column is trapped in cooled charcoal. At minute 100, xenon begins to exit the charcoal column, and the flow is redirected to a condenser, where the xenon is collected. The black dashed line is a theoretical expectation from Eq.~\ref{eq:ElutionCurve}, applied only to the krypton behavior, which is in the appropriate linear regime. This cycle used 2~kg of xenon, and helium as a carrier gas.}
\label{fg:gasSeparation}
\end{figure}

Adsorption-based chromatography utilizes the difference of the adsorption strength among the adsorbates. In a chromatographic column, a ``carrier fluid" flows through the bed of the adsorbent, carrying along a dilute adsorbate fluid. The average speed of each adsorbate through the bed is dependent on the fraction of time it spends in the mobile phase. Adsorbates that bond more strongly to the adsorbent spend smaller fractions of their time in the mobile phase and thus have lower average speed through the column. The carrier fluid itself is chosen to rarely interact with the adsorbent. Fig.~\ref{fg:gasSeparation} shows xenon and krypton exiting a charcoal column at different times. 

\begin{figure}
\centering
\includegraphics[width=0.5\textwidth]{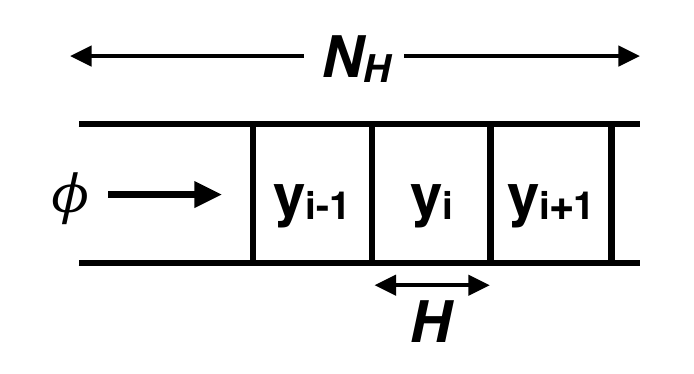}
\caption[Chromatography column]{A chromatographic column is approximated as a series of $N_H$ height-equivalent theoretical stages (HETS). Each stage contains a charcoal mass $m$ in a volume $V$ and an height of $H$. The mobile phase moves through the column with a volume speed of $\phi$. The number density of the adsorbate in the mobile phase at the $i^{th}$ stage, $y_i$, depends on $\phi$ and $y_{i-1}$.}
\label{fg:columnAnalysis}
\end{figure}

A commonly used simplified model of the propagation of an adsorbate in a chromatographic column assumes that the chromatographic column consists of $N_H$ height-equivalent theoretical stages (HETS) as shown in Fig.~\ref{fg:columnAnalysis}. Each HETS has a volume $V$ and contains adsorbent of mass $m$. The number of gas molecules in each stage is the sum of those in the gaseous and adsorbed phases, and can be written using the terms defined in Eq.~\ref{eq:adsorptionConstant}:
\begin{equation}
N = Sm+yV \approx Sm = ykm. \label{eq:firstPrinciple}	
\end{equation}
The approximation holds when the number of molecules adsorbed is much greater than the number in the gas phase. 

In the $i^{th}$ stage, the change in the number of molecules in time $dt$ is the difference between the incoming flux and the outgoing flux: 
\[dN_{i}=y_{i-1}\phi dt-y_{i}\phi dt.\]
where $\phi$ is the volume swept by unit time, or volume flow. 

A simple differential equation can be written for $y_i$: 
\begin{equation}
\frac{dy_{i}}{dt}=-\frac{N_H}{\tau}(y_{i}-y_{i-1}),
\label{eq:ElutionDiffEq}
\end{equation}
where
\begin{equation}
\tau \equiv \frac{k M}{\phi}
\label{eq:RetentionTime} 	
\end{equation}
is referred to as ``retention time." Here, $M \equiv N_H m$ is the total mass of the adsorbent in the column. 

An analytic solution of Eq.~\ref{eq:ElutionDiffEq} exists for a special case when the column is initially empty of adsorbates and the input feed of the adsorbates looks like a Dirac-delta function at $t=0$:
\begin{equation}
y(t)=\frac{N_H^{N_H}}{\Gamma(N_H)}{\left(\frac{t}{\tau}\right)}^{N_H-1} e^{-N_H\frac{t}{\tau}},\label{eq:ElutionCurve}
\end{equation}
where $\Gamma$ is gamma function. Eq.~\ref{eq:ElutionCurve} is referred to as an elution curve, and its integral from zero to asymptote is unity. It has a global maximum at $t = \frac{N_H-1}{N_H}\tau$. For large $N_H$, this is very close to $\tau$, which is also called the ``breakthrough time". For faster production and finer separation of gases, lower $\tau$ and higher $N_H$ are desired. 

The height $H$ of a HETS depends on the linear velocity $u$ of the mobile phase in the column as parametrized by Van Deemter~\cite{VanDeemter}:
\begin{equation}
H = A + \frac{B}{u} + C u\label{eq:VanDeemter}.
\end{equation}
$A$ represents the contribution from Eddy currents in the column, \textit{i.e.}, the many possible paths the molecules can find in the packed column. $B/u$ represents diffusion in the longitudinal direction; and $Cu$ represents the dispersion due to the non-uniformity of $u$, mostly caused by the porosity of the stationary phase. Eq.~\ref{eq:VanDeemter} provides guidance as to the optimal flow rate of the mobile phase and the optimal shape of the column. The minimum $H$, which maximizes $N_H$, occurs at $u=\sqrt{\frac{B}{C}}$. A modern summary of the Van Deemter equation and its coefficients for gas chromatography can be found in Ref.~\cite{Engewald:2014jr}.

Eq.~\ref{eq:ElutionCurve} assumes that the density of the adsorbates in the adsorbed state is linearly proportional to its partial pressure as shown in Eq.~\ref{eq:henry}, and it does not count the competing adsorption between the multiple adsorbates. Because the goal of our production system is to process the greatest mass of xenon per unit time, a high ratio of adsorbate molecules to adsorbent mass is preferred, and the observed output deviates from Eq.~\ref{eq:ElutionCurve}. The deviation leads to earlier breakthrough of xenon and the broadening of the peaks, both of which result in worse separation. Saturation of the adsorbent sites with xenon limits the mass of xenon that can be processed in a cycle as we optimize for fast production at adequate separation. 

Although chemically inactive, noble gases such as xenon and krypton can adsorb on activated charcoal or molecular sieve~\cite{Munakata99, Bazan:2011ec}. Polarization of their electron shells by the induced electric dipoles of the charcoal surface leads to an attractive potential. Its strong affinity can be explained by the large conductive microscopic surface area. Activated charcoal is readily available commercially.

Helium is a good mobile phase because it is chemically extremely stable and does not compete for adsorption on the charcoal~\cite{MoellerUnderhill}. Helium does not have any naturally occurring radioactive isotopes, and can be easily removed by freezing the xenon on a liquid-nitrogen cooled surface and pumping away the helium carrier gas.

\section{System Design \& Operation}

\begin{figure}
\includegraphics[width=1\textwidth]{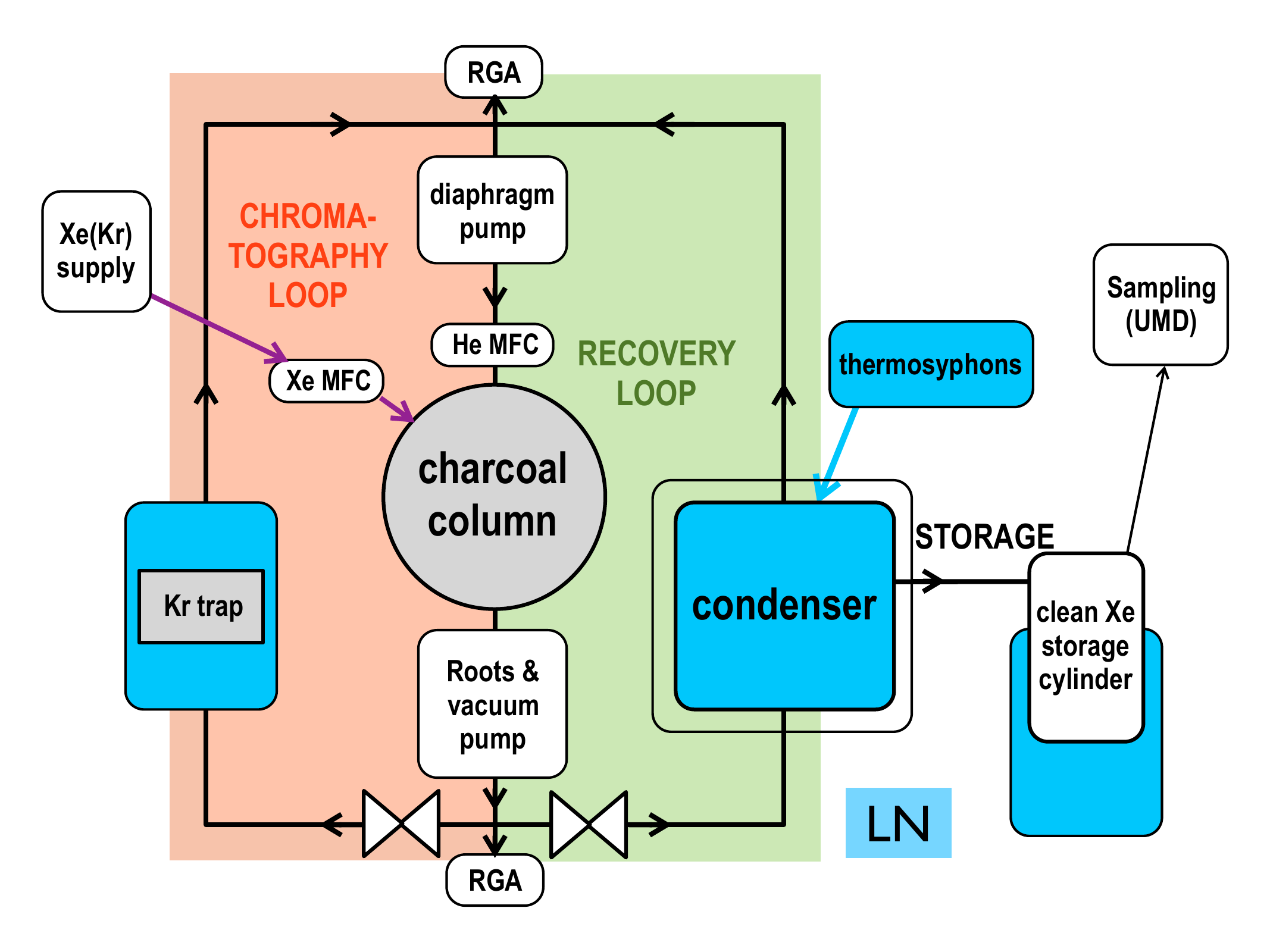}
\caption[Kr Removal System Block Diagram]{Block diagram of the LUX Kr removal system at Case Western Reserve University~(CWRU). The xenon supply with its trace krypton enters the charcoal column under the influence of a circulating stream of helium. The helium first carries out the krypton, which is collected in a charcoal trap at 77~K. Then the xenon is detected exiting the column and the valves are set to direct the column output to the condenser, where the xenon is collected at 77~K. After multiple such cycles, the helium is pumped away and discarded, and the processed frozen xenon is warmed and cryo-pumped into a storage cylinder for transportation. While the xenon is cryo-pumped, the krypton trap is separately warmed and purged with helium to clean it for the next round.}\label{fg:FlowState}
\end{figure}

A schematic diagram of the LUX krypton removal system is shown in Fig.~\ref{fg:FlowState}. More technical details are available in~\cite{mythesis}. A regulated flow of xenon containing a trace contamination of krypton is fed into a 60~cm $\times$ 60~cm (diameter $\times$ height) column filled with 60~kg of activated OVC 4x8 charcoal from CalgonCarbon$\textsuperscript{\textregistered}$ at ambient temperature. The xenon is injected into the top of the charcoal column while the helium carrier gas continues to circulate through the column. The carrier gas moves the krypton and xenon through the column at different rates due to the difference in their adsorption strength. The krypton exits the column first, and flows into a ``krypton trap." The krypton trap is a 1.5-inch-diameter stainless steel tube filled with 500~g of activated charcoal that is immersed in a liquid nitrogen bath~(77~K). At this temperature the krypton retention time is much longer than the processing time, and the krypton effectively freezes on to the charcoal while the helium carrier gas is still highly mobile. The purified helium gas exits the trap and circulates back into the column. This process is called the ``chromatography cycle," and lasts until xenon emerges out of the column. During the chromatography cycle, the mass flow of helium is driven by a diaphragm pump and regulated by a mass flow controller in front of the column to stabilize the flow rate and the pressure inside the charcoal column. 

The ``xenon recovery cycle" is triggered when xenon emerging from the column is detected by a sampling residual gas analyzer (RGA). The xenon-helium mixture is redirected from the charcoal column into a condenser. The condenser is a cryogenic vessel with an internal volume of about 1~m$^3$, cooled to 77~K. While the processed xenon is frozen inside the condenser, helium gas passes through unimpeded and is fed back into the charcoal column to complete the cycle. A Roots blower (Edwards EH250) connected in series with its backing pump (Alcatel Adixen ACP40) at the output of the column generate 2000~liter per minute (lpm) of volume flow to accelerate the xenon recovery. The pressure inside the column is kept at 10--20~mbar. The RGA at the output of the backing pump, whose details are discussed in Sec.~\ref{sec:sampleassay}, monitors the output gas from the column in real time. Once all the processed xenon is recovered from the column, a new cycle with the next batch of raw xenon begins. After several cycles, the helium carrier is pumped out of the condenser and discarded, and the accumulated clean xenon is warmed, evaporated, and transferred into a storage cylinder for transportation to the LUX experiment. 

The system is designed to mitigate the possible sources of contamination. One such source is air: 1 liter of air contains as much krypton as the final 400~kg of xenon after purification to the 4~ppt LUX goal. The system is vacuum-sealed to minimize contamination from external air. Another concern is that krypton can dissolve in plastic components or pump lubricant and can be released at a later time, mixing with the purified xenon. To minimize such cross-contamination, most of the system is made of stainless-steel tubing and most of its joints are sealed by metal gaskets. There still are a few non-metallic components that cannot be removed, such as rubber O-rings and filters. A rough estimate of the cross-contamination through these components is less than $1\times10^{-7}$ of the total krypton~\cite{mythesis}.

The separation and the production rates depend on the pressure and the volume flow rate inside the charcoal column. For our application, we wish to maximize the mass of xenon processed per batch, and this requirement forces us to operate in the partially-saturated non-linear regime, where the elution curve estimation from Eq.~\ref{eq:ElutionCurve} no longer holds. The optimal operation parameters for the maximum process rate and the krypton reduction factor were found empirically by measuring the output concentration under various conditions. The raw xenon begins with a krypton contamination at a concentration of 130~ppb, below the threshold directly visible with a commercial sampling RGA. To visualize the time-dependent concentration of the krypton exiting the charcoal column, a 1:100 mol/mol mixture of krypton and xenon is used. Fig~\ref{fg:gasSeparation} shows the mixture separated in the system monitored by the RGA. Measurements from different column pressures, xenon feed rate, and xenon feed mass were compared. 

For the chromatography cycle, maximal separation of the peaks is the primary concern. The pressure inside the charcoal column is directly related to the diffusion constant of the gas, represented by the $B$ term in Eq.~\ref{eq:VanDeemter}. In the system, the cross-sectional area of the column is larger than the optimal area based on the pumping speed, and the system operated in the region where the $B$ term in Eq.~\ref{eq:VanDeemter} dominated. The results also supported the expectation that a higher pressure inside the column leads to larger separation of the peaks and a narrower krypton peak. We find that a 500~mbar column pressure with a 50~standard liter per minute~(slpm) helium flow rate are the optimal chromatography conditions for our system.

The primary concern for the xenon recovery cycle is its duration. Due to its strong bonding, xenon moves slowly through the charcoal column, and the duration of the xenon recovery cycle dominates the overall production rate. Eq.~\ref{eq:RetentionTime} suggests that the xenon retention time can be reduced by using a higher volume flow rate. The Roots blower provided a five-fold higher volume flow rate compared to the system described in~\cite{ABolozKr} at pressures in 5--30~mbar range, and kept the recovery cycle duration at three hours even though the charcoal column mass had increased sixfold.

Once the system was built, the optimal operating parameters were found through a series of tests. The feed rate and the feed mass of xenon can change the quality of the separation. While we wish to maximize the production mass per unit time, too much xenon relative to charcoal leads to non-optimal saturation of the adsorbent and worse separation. Similarly, a high xenon feed rate can locally saturate charcoal and lead to worse separation. Again, because the adsorption is nonlinear, the optimal feed rate was found empirically. 

A series of tests using the enriched krypton mixture leads to the following parameters for chromatography in the system. The column is kept at a pressure of 500~mbar during chromatography, with a 50~slpm helium mass flow. A 2-kg batch of xenon is injected over 20~minutes. The krypton output peaks at about 40~minutes after the start, and lasts until xenon emerges from the column, at about 115~minutes. The detection of xenon triggers the recovery cycle. First, the output flow is re-directed to the condenser, and the vacuum backup pump reduces the column pressure below 100~mbar. When the pressure is low enough, the Roots blower activates and recovers the clean xenon from the column into the condenser. A constant helium flow of 15--20~slpm keeps the pressure inside the column at 10--20~mbar. The recovery cycle lasts three hours, and one full cycle processes 2~kg of xenon in 5~hours. A typical source cylinder contains about 50~kg of xenon, which is processed in 25 cycles. The processed xenon is accumulated in the condenser. When the source cylinder is empty, the helium in the condenser is pumped out, and the processed xenon is warmed and transferred via pressure gradient into a storage cylinder that is cooled by liquid nitrogen.

A total 395~kg of xenon was processed for LUX in four months, between the 15th of September 2012 and the 10th of January 2013. Most of the processes were executed automatically with minimal operator intervention. The process of pumping away the helium and transferring the xenon to a storage bottle was done manually, and typically took one day. Less than 1~kg (or 0.2\%) of xenon was lost, mostly due to operator errors. The key parameters are summarized in Table~\ref{tb:krRemovalComparison}, and compared with the results from other systems.

\begin{table}
\caption{Comparison of Kr removal techniques.}
\begin{center}
\begin{tabular}{c | c | p{3cm} | p{3cm} | c}
 & method & $^{85}$Kr concentration (best g/g)  & $^{85}$Kr concentration in experiment (g/g)  &process speed (kg/h)\\
\hline
this work & chromatography & 4~ppt & $<0.2$~ppt\footnotemark[3]  & 0.4\\
XMASS~\citep{XMASSKr}  & distillation & 2~ppt & 2.1~ppt & 5\\
XENON100~\citep{Lindemann:2014ea} & distillation & 0.6~ppt & 0.6~ppt & unknown\\
Panda-X~\citep{Wang:2014dt}  & distillation & 13~ppt & 28~ppt~\citep{Tan:2016fv} & 5\\
XENON1T~\citep{Fieguth:2016wbj} & distillation & reduction $>10^5$ & unknown & 3\\
\end{tabular}\label{tb:krRemovalComparison}
\end{center}
\end{table}
\footnotetext[3]{processed twice}

\section{Product Sampling \& Assaying}\label{sec:sampleassay}

Two sampling RGAs are used to monitor the composition of the gas in the system in real time. One RGA is mounted after the output of the vacuum pump to monitor the gas exiting the charcoal column before it enters the Kr trap or the condenser. The second RGA monitors the gas outputs of the Kr trap and the condenser. 

In addition, at every transfer from the condenser to the storage cylinder, a sample of the processed xenon is collected from the storage cylinder. The sample reflects the average gas concentrations from the multiple cycles that went into the storage cylinder.

The processed xenon from a single cycle can be sampled, too. A four-liter evacuated cryogenic bottle is attached to a spur from the path from the vacuum pump to the condenser. The valve to the bottle opens to the xenon stream during the xenon recovery phase, and the sample is collected until the pressure of the system is equalized by the helium in the stream. 

\begin{figure}[h!]\centering
\includegraphics[width=\textwidth]{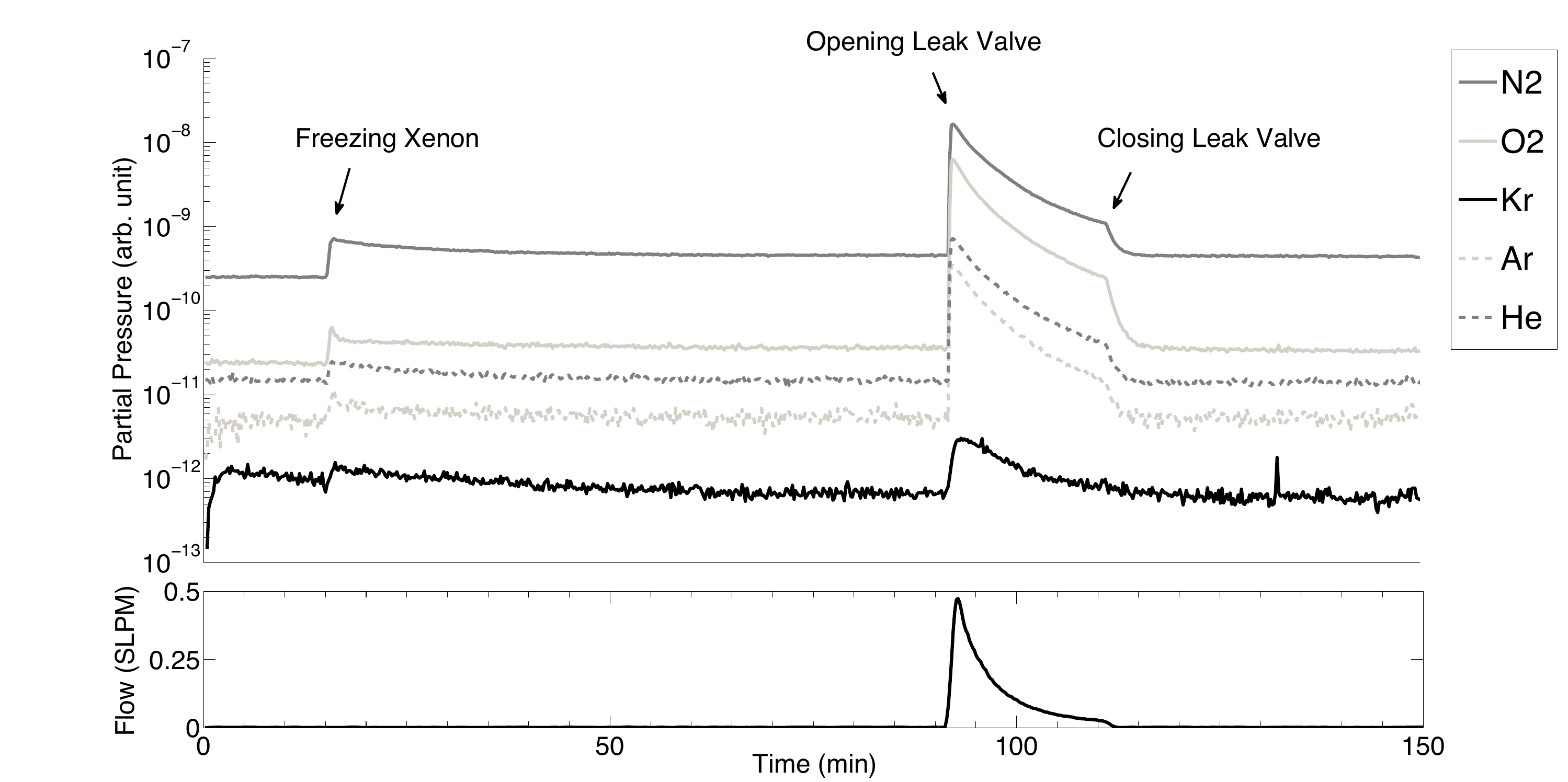}
\caption{(top) A xenon sample (LSB3) in which an 18.2 ppt krypton signal is detected. Each line corresponds to the partial pressure of a gas in the sample. The initial rise in partial pressures is due to freezing a small amount of xenon in the cold trap, while the later rise and fall in the partial pressures is due to opening and closing a leak valve to begin and end the assay process. (bottom) The flow rate into the cold trap is indicated.}
\label{fig:assaying}
\end{figure}

The krypton content of the processed xenon samples is measured independently off site using a high-sensitivity assaying system at the University of Maryland. The system utilizes a liquid nitrogen cold trap to separate impurities from the xenon. When the xenon sample flows through the cold trap, the bulk xenon freezes as it contacts the surfaces cooled by a liquid nitrogen bath, while impurities such as krypton mostly pass through in observable quantities. The changing concentration of the impurities over the process of the assaying is shown in Fig.~\ref{fig:assaying}. The absolute level of impurities is deduced by comparing the krypton to calibration samples. The sensitivity to krypton is 0.3~ppt~\cite{UMD2010,UMD2011}. 

\begin{figure}[h!]\centering
\includegraphics[width=\textwidth]{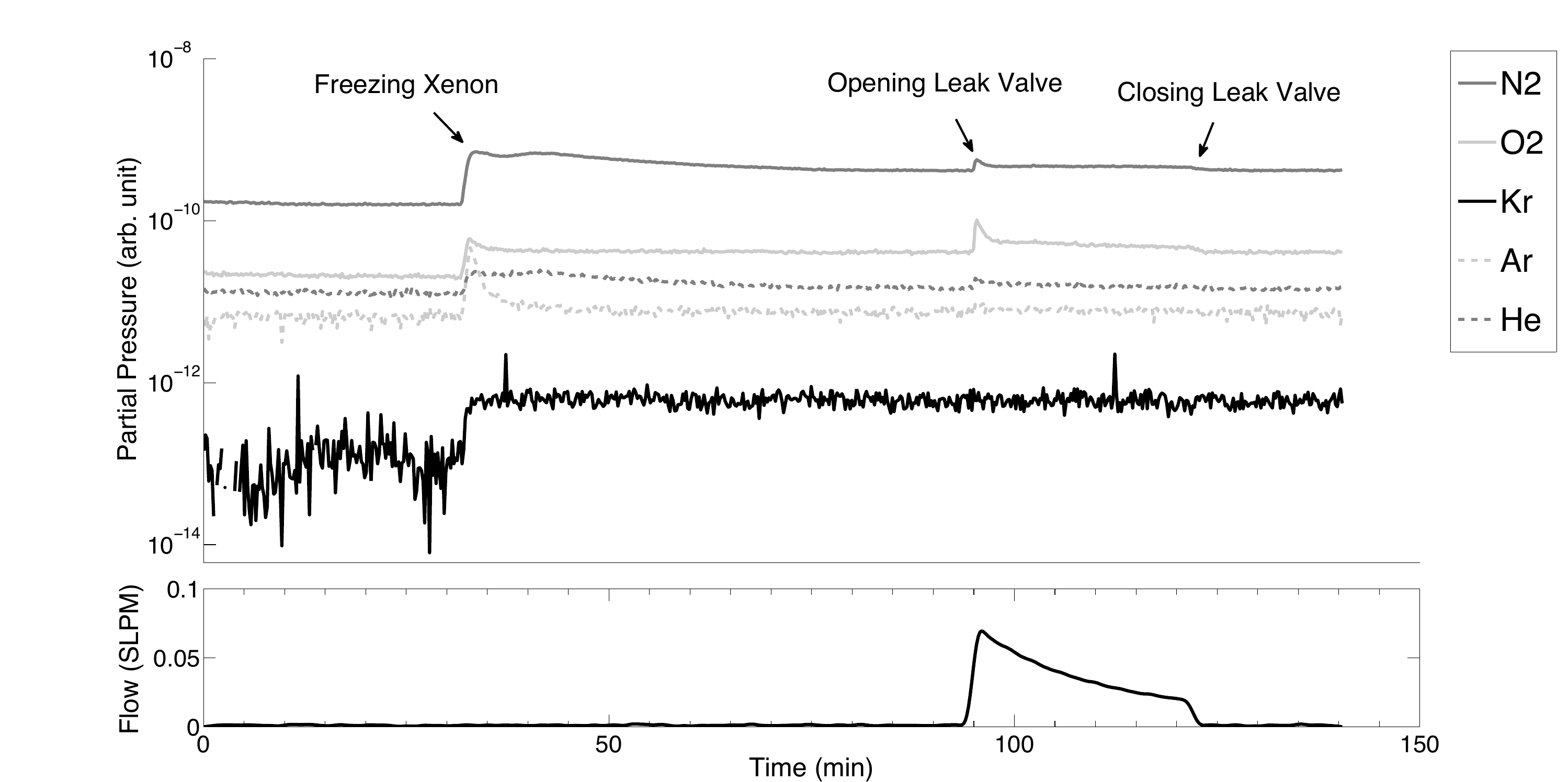}
\caption{A xenon sample (LSB7) in which a krypton signal is not detected. The flat krypton partial pressure is shown in bold solid, and the scaled flow rate is shown in black. No krypton was detected in this sample and set a limit of 40~ppt based on the signal noise and flow rate during the sample. The same sample was later remeasured with higher sensitivity, and 4~ppt of krypton was found.} 
\label{fig:noKrSignal}
\end{figure}

We claim no krypton detection when the krypton partial pressure does not trace the flow rate over time as shown in Fig.~\ref{fig:noKrSignal}. An upper limit is set by assuming that a signal one standard deviation above the noise can be detected. Higher flow rates produce larger signals and allow for the best sensitivity.

\begin{figure}
\centering
\includegraphics[width=\textwidth]{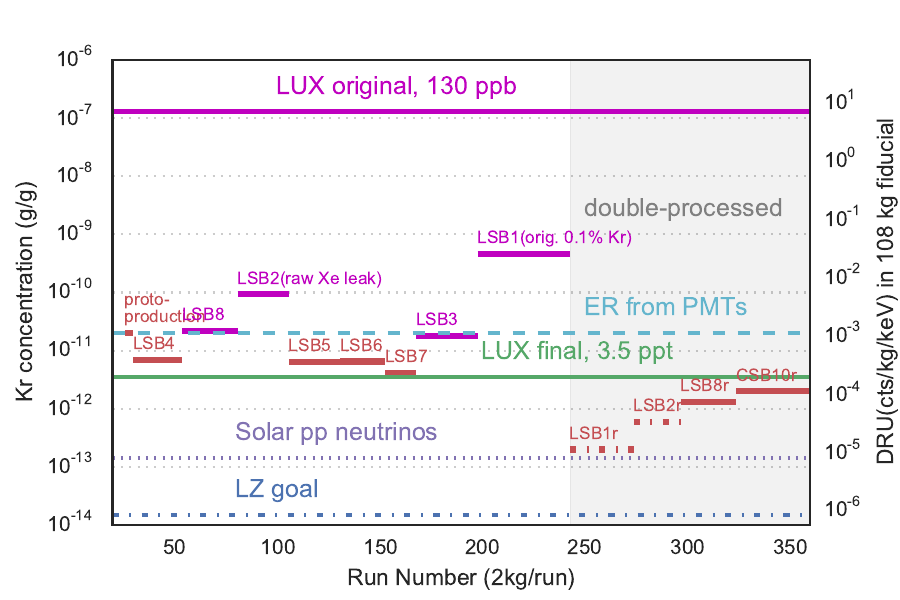}
\caption[Krypton Removal Progress]{Progress of krypton removal runs for 395~kg of xenon for LUX. The $x$-axis represents the run numbers, each corresponding to 2~kg of xenon, in chronological order. The $y$-axes represents the krypton level in two different scales; the mass concentration (left), and the radioactivity rate (right). The xenon procured for LUX initially contained 130\,ppb of krypton. The production reduced the average contamination down to 4\,ppt. Runs after 243 are the reprocessing of the batches that are marked with the magenta lines and had  a krypton concentration higher than our target value. They are marked with an "r" at the end of their indices. The system was cleaned before Run 243 to minimize cross-contamination from the trace amount of krypton accumulated in the system. The thick colored lines refer to the LUX Storage Bottle batch number, each corresponding to about 50\,kg. The krypton level marked with the dot-dashed lines indicate upper bounds for batches with no detected krypton in the assay.}
\label{fg:krremovalproduction}
\end{figure}

The assay results from the production run are summarized in Fig.~\ref{fg:krremovalproduction}. The average concentration of krypton dropped from 130~ppb to 4~ppt. The average reduction factor is $3\times10^4$, including batches that were processed twice. The best reduction factor from a single processing is also about $3\times10^4$. One of the double-processed batches had $< 0.2$~ppt of krypton, a limit set by the assaying sensitivity. More tests are necessary to determine whether fundamental chromatography or cross-contamination limited the reduction factor.

In addition to krypton, the assay measured the levels of other impurities including nitrogen (57~ppb), oxygen (16~ppb), argon (1.3~ppb), and methane ($<1$~ppb). All are lower after the production. Residual helium from the processing is a concern because it can degrade the PMTs by diffusing through the quartz windows. Assays indicated that the helium concentration was reduced to 3.2~ppb, far below that in air, and it presented no threat.  

Analysis of the WIMP search data from the 2013 run of the LUX experiment independently constrained the $^{85}$Kr contamination in the xenon by searching for its decay signature. A small fraction (0.434\%) of $^{85}$Kr decays can be tagged by their unique signatures: a 173~keV beta followed by a 514~keV gamma from de-excitation of a $^{85}$Rb metastable state with 1.015$\mu$s half-life~\cite{Firestone:471274}. Analysis of the  data saw no such events, and set a 90\% confidence upper limit of $<0.26$~mDRU, or $<5.4$~ppt Kr content. If we assume the 3.5\,ppt krypton concentration from the assay above, the number can be interpreted as an upper limit of 31~ppt $^{85}$Kr/Kr ratio in the atmosphere, consistent with the expected upper limit of atmospheric concentration of 20~ppt. 

\section{Conclusion}
An adsorption-based gas-charcoal chromatographic system was built at the Case Western Reserve University and used to reduce the krypton concentration in the LUX target xenon. The processed xenon contained 3.5\,ppt of krypton, surpassing the LUX goal. The average reduction factor was $3\times10^4$, and about the same reduction was achieved from a single pass. The best batch from the double-production contained less than 0.2~ppt, the measurement sensitivity. The chromatography system was capable of processing and storing 50\,kg of xenon a week with minimal human intervention. The processed xenon has been used for the scientific runs of the LUX experiment~\cite{Akerib:2014uk}.

A krypton removal system that can produce lower krypton levels with higher production rate is required for larger xenon-based dark matter experiments such as the LUX-ZEPLIN~(LZ) experiment. LZ is a scaled-up successor to LUX, planned to operate in 2020~\cite{Akerib:2015wr}. It is designed to reach a WIMP-nucleon cross section sensitivity of $2\times10^{-48}$~cm$^2$ with 7~tonnes of active xenon mass from a 10-tonnes total xenon mass. An irreducible electron-recoil background is set by elastic scattering of solar $pp$ neutrinos, whose rate is comparable to 0.2~ppt krypton dissolved in xenon. The LZ collaboration aims to reduce its krypton concentration below 0.015\,ppt. Further investigations are ongoing to determine the ultimate floor that can be reached by improving the design to further reduce cross-contamination and air ingress, which can otherwise compromise the reductions allowed by repeated processing. If the target is met, LZ is expected to directly measure the scattering of solar $pp$ neutrinos on electrons and coherent elastic scattering of $^8$B neutrinos on nuclei, in addition to substantially extending the sensitivity to WIMP dark matter interactions.

\section*{Acknowledgements}
This work was partially supported by the U.S. Department of Energy (DOE) under award numbers DE-FG02-08ER41549, DE-FG02-91ER40688, DE-FG02-95ER40917, DE-FG02-91ER40674, DE- NA0000979, DE-FG02-11ER41738, DE-SC0006605, DE-AC02-05CH11231, DE-AC52-07NA27344, and DE-FG01-91ER40618; the U.S. National Science Foundation under award numbers PHYS-0750671, PHY-0801536, PHY-1004661, PHY-1102470, PHY-1003660, PHY-1312561, PHY-1347449, PHY-1505868; the Research Corporation grant RA0350; the Center for Ultra-low Background Experiments in the Dakotas (CUBED); and the South Dakota School of Mines and Technology (SDSMT). LIP-Coimbra acknowledges funding from Funda\c{c}\~{a}o para a Ci\^{e}ncia e Tecnologia (FCT) through the project grant PTDC/FIS-NUC/1525/2014. Imperial College and Brown University thank the UK Royal Society for travel funds under the International Exchange Scheme (IE120804). The UK groups acknowledge institutional support from Imperial College London, University College London and Edinburgh University, and from the Science \& Technology Facilities Council for PhD studentships ST/K502042/1 (AB), ST/K502406/1 (SS) and ST/M503538/1 (KY). The University of Edinburgh is a charitable body, registered in Scotland, with registration number SC005336.

This research was conducted using computational resources and services at the Center for Computation and Visualization, Brown University.

We gratefully acknowledge the logistical and technical support and the access to laboratory infrastructure provided to us by the Sanford Underground Research Facility (SURF) and its personnel at Lead, South Dakota. SURF was developed by the South Dakota Science and Technology Authority, with an important philanthropic donation from T. Denny Sanford, and is operated by the Lawrence Berkeley National Laboratory for the Department of Energy, Office of High Energy Physics.

\bibliography{KrRemovalLibrary}

\end{document}